# Deep Learning Driven Enhancement of Optical Vortex Line Robustness in Atmospheric Turbulence

Dmitrii Tsvetkov*, Danilo Gomes Pires, and Natalia Litchinitser

*Department of Electrical and Computer Engineering, Duke University, Durham, North Carolina 27708, USA*

*\*dmitrii.tsvetkov@duke.edu*

**Abstract**

The stability of optical vortex structures in turbulent environments is critical for their applications in optical communication, quantum information, and structured light technologies. Although topological invariants, such as crossings and linking numbers, are fundamentally invariant, recent studies reveal that their observed values deteriorate considerably in turbulent conditions due to environmental effects. In this study, we introduce an alternative approach based on the geometric stability of three-dimensional singularity line shapes, demonstrating that shape-based tracing of singularities outperforms both topological and spectral methods in turbulence. To test this concept, we propose Flower Beams, a novel class of structured optical fields featuring controllable petal-like singularity morphologies. We construct an 81-element optical alphabet and classify these structures after turbulence using deep learning. Our findings reveal that shape-based tracing achieves classification accuracy exceeding 90% in the weaker turbulence regimes and remains highly competitive even in stronger turbulence, significantly outperforming spectral and topology-based approaches. Experimental results confirm that the predicted shape stability holds in real-world conditions. This study establishes the shape of the singularities' lines as a scalable and resilient alternative for structured light tracing and transmission, opening new avenues for turbulence-robust applications.

**Introduction**

Optical singularities play a fundamental role in the science of structured light, with applications spanning optical trapping [1, 2], imaging [3], quantum information [4–6], metrology [7, 8], and telecommunications [9, 10]. Among these singularities, phase singularity lines, corresponding to zero light intensity and undefined phase, are fascinating due to their complex evolution in space and the promise of applications ranging from fundamental physics to advanced optical systems [11,12]. The ability to control and manipulate these singularity lines enables the generation of intricate three-dimensional field structures, including vortex knots [13–16], which have been studied in various fields, ranging from condensed matter systems and fluid dynamics to photonics [13,15–29]. Optical vortex knots are three-dimensional (3D) singularities in light beams that form closed-loop lines of darkness surrounded by light. Mathematically, knots are topologically protected structures, characterized by invariants such as linking numbers and polynomials, which remain unchanged under continuous deformations [30,31]. While these properties suggest that optical knots should exhibit robustness against environmental perturbations [32], recent studies have shown that the number of crossings (a fundamental topological invariant) remains constant in weak turbulence but changes in stronger turbulence conditions [33,34].

Several approaches have been developed to enhance the stability of knots in complex environments, including beam and singularity shaping, as well as alternative detection strategies [13, 35. Nonetheless, findings indicate that mathematical topological protection does not always translate directly to physical stability in optical systems [33, 34, 41]. Given the potential of optical knots as information carriers for free-space optical and quantum communication [42–45], quantum computing [46,47], and microfabrication [48,49], there is a critical need to investigate alternative approaches to knot representation that may offer greater stability than the existing topological invariants. In this work, we compare two key strategies: (i) assessing the stability of multiplexed Laguerre-Gaussian (LG) mode superposition comprising the knots and (ii) analyzing the structural resilience of the 3D geometries of the phase singularities in turbulence. To evaluate these strategies, we define two optical "alphabets" [50,51]: alphabet-11, which consists of 11 distinct optical knots, 9 Hopf links, and 2 Trefoil knots of different shape (see Section 2 for details) – and alphabet-81, comprising 81 distinct examples of a novel class of beams that we named Flower Beams (see Section 6). Using both the Mean Squared Error (MSE) approach and deep neural networks (DNNs), we theoretically and experimentally demonstrate that the 3D shape of vortex lines proves considerably more stable under perturbations than standard topological invariants, surpassing the stability observed in mode spectrum-based methods.

**Results**

**Analysis of Optical Knots in Turbulent Conditions**

Turbulence is a complex and highly dynamic phenomenon arising from irregular fluid or gas flow, resulting in chaotic fluctuations in pressure, velocity, and density [52–54]. In the optical context, turbulence mainly arises from refractive index variations in the propagation medium (*e.g.*, atmosphere or underwater) [55,56]. These variations affect both the phase and amplitude of light waves, introducing wavefront aberrations, beam wandering, and intensity scintillations [56]. Such effects have a significant impact on structured light applications, including optical communications and high-resolution imaging [57,58].

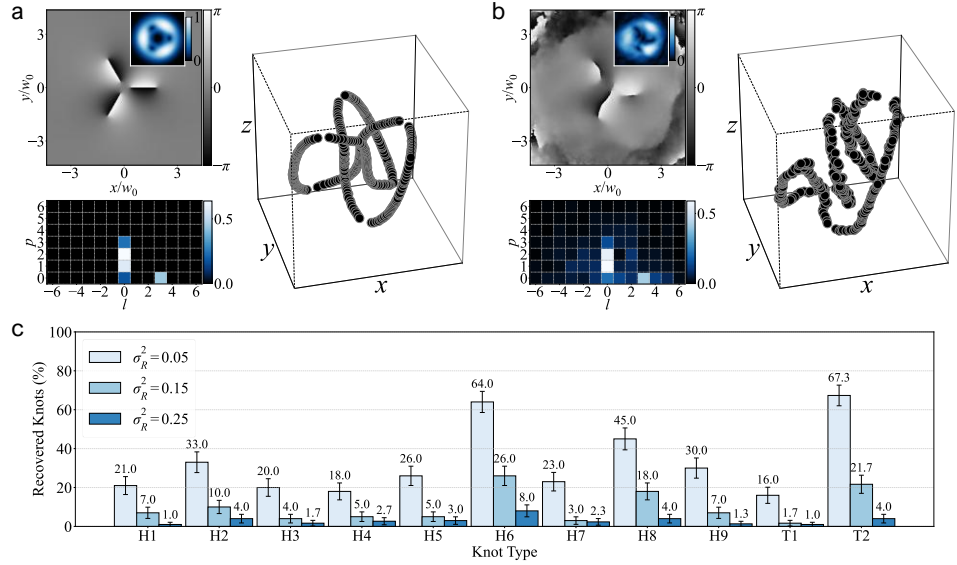

**Fig. 1. The impact of turbulence on trefoil knot T1 alongside a statistical analysis of recovery rates for the Alphabet-11 knots.** (a) Trefoil knot T1 in an unperturbed environment. The top-left panel displays the phase of the electric field in the central plane (z = 0), with an inset showing the corresponding amplitude distribution. The bottom-left panel presents the LG spectrum of the knot, and the right panel offers a 3D visualization of its singularity line structure. Panel (b) uses the same arrangement as panel (a) to display trefoil knot T1 after it has experienced turbulence of strength $\sigma_R^2 = 0.15$, highlighting the changes in phase, LG spectrum, and singularity structure induced by the turbulent environment. Panel (c) features a histogram that quantifies the percentage of knots recovered at three different turbulence strengths ($\sigma_R^2 = 0.05, 0.15,$ and $0.25$) across the complete Alphabet-11 knot set (H1, H2, …, H9, T1, T2).

Although optical knots were initially expected to exhibit the same robustness as their mathematical counterparts, studies have shown that optical knots and related topological structures are vulnerable to medium- and strong-turbulence [38, 59]. An optical knot can be constructed as a coherent superposition of weighted LG modes, and the "spectrum" refers to the distribution of modal weights across radial and azimuthal indices. Turbulence induces spatial and temporal variations in the refractive index, resulting in localized distortions in phase and intensity distributions (Fig. 1a, b, top left) and broadening of the LG mode spectrum (Fig. 1a, b, bottom left). Beyond intensity fluctuations and LG spectrum broadening, turbulence also deforms phase singularity lines within a vortex field (Fig. 1a, b, right). Even moderate turbulence can introduce significant distortions in optical knots, ultimately degrading their topological integrity. Under strong turbulence, singularity lines may become highly fragmented, effectively destroying the knotted structure and severely limiting the practical viability of such fields [21,34,62]. Preserving the integrity of optical knots in turbulent conditions remains a significant challenge, driving the search for more turbulence-resilient structured beams.

We begin by analyzing the robustness of optical knots in turbulence, assessing spectral stability in the LG basis, and topological persistence, which is defined by the preservation of the knot topology [34]. We consider a set of 11 optical knots, Alphabet-11, which includes nine Hopf links (H1–H9) and two trefoil knots (T1 and T2) with varying geometries (see Supplementary Materials S1 for details on the basis generation). Following our previous approach to assessing optical knot stability in turbulence [34], we first evaluate the topological robustness of each knot in Alphabet-11 after its exposure to turbulence. In this context, topological robustness refers to a knot's ability to preserve its original singularities' crossings and linking structure despite perturbations. Figure 1a shows a three-lobed structure of the original trefoil knot T1 with three crossings. However, after exposure to turbulence (Fig. 1b), a reconnection event disrupts the central crossings, unraveling the knot and transforming it into an unknotted loop, corresponding to the loss of topological information. Additional examples of knots subjected to different turbulence strengths are provided in Supplementary Materials S2.

To explore the impact of turbulence, we simulated three distinct turbulence regimes of varying strengths, characterized by the Rytov variance parameter $\sigma_R^2$ (see Methods for details). Here we consider three levels of turbulence strength, including the weakest turbulence $\sigma_R^2 = 0.05$, a moderate strength with $\sigma_R^2 = 0.15$, and the strongest turbulence regime with $\sigma_R^2 = 0.25$. For each optical knot in Alphabet-11, we analyzed 300 turbulence realizations per turbulence strength, assessing whether the topology (i.e., the number of crossings) remained intact. Figure 1c summarizes the results, showing the percentage of knots that retained their original topology. Even the most resilient knots (e.g., H6, T2, both optimized for turbulence [34]) show a significant loss of stability, with moderate turbulence reducing their persistence below 26%, and the strongest turbulence further decreasing it to 8% (H6) and 4% (T2). Other knots exhibit even lower stability rates. Across all turbulence strengths, the average correct knot recovery rate for Alphabet-11 is 15.3%, with individual rates of 33.0% in weak turbulence, 9.8% in moderate turbulence, and just 3.0% in the strongest turbulence.

These results underscore the limitations of topology-based tracing in turbulent environments and emphasize the need for more robust approaches if optical knots are to be effectively utilized in free-space propagation-based applications under perturbative conditions. To further understand the limitations of topology-based tracing and explore alternative approaches, we assess the stability of the LG spectrum compositions that define each knot and analyze the resilience of their 3D singularity line structures in turbulence. We employ two complementary approaches: (i) a mathematical framework using MSE analysis to quantify spectral distortions and(ii) a machine learning-based classification method to evaluate the robustness of optical knots and Flower Beams in turbulent conditions.

**Mean Square Error Analysis of LG Spectra of Alphabet-11**

To evaluate the stability of Alphabet-11 optical knots in turbulence, we employ a mathematical approach based on MSE analysis, focusing on both LG spectra and singularity line shapes. MSE provides a simple yet powerful metric to quantify the difference between two distributions by averaging the squared deviations of corresponding components. We begin by analyzing the MSE of optical LG spectra to quantify spectral distortions caused by turbulence. Specifically, we compare the post-turbulence LG spectra of each optical knot to its pre-turbulence basis spectra using Full-Spectrum MSE and Filtered-Spectrum MSE. The first method calculates the MSE between each LG mode weight in the post-turbulence spectra and the original basis spectra, considering radial indices $p$ from 0 to 10 and azimuthal indices $l$ from -10 to 10. This approach provides a comprehensive assessment of spectral distortion across the entire mode space. The Filtered-Spectrum MSE method computes the MSE only for modes that were present in the original LG spectrum before turbulence. By focusing on dominant modes, this approach isolates specific spectral changes while reducing noise from newly introduced modes.

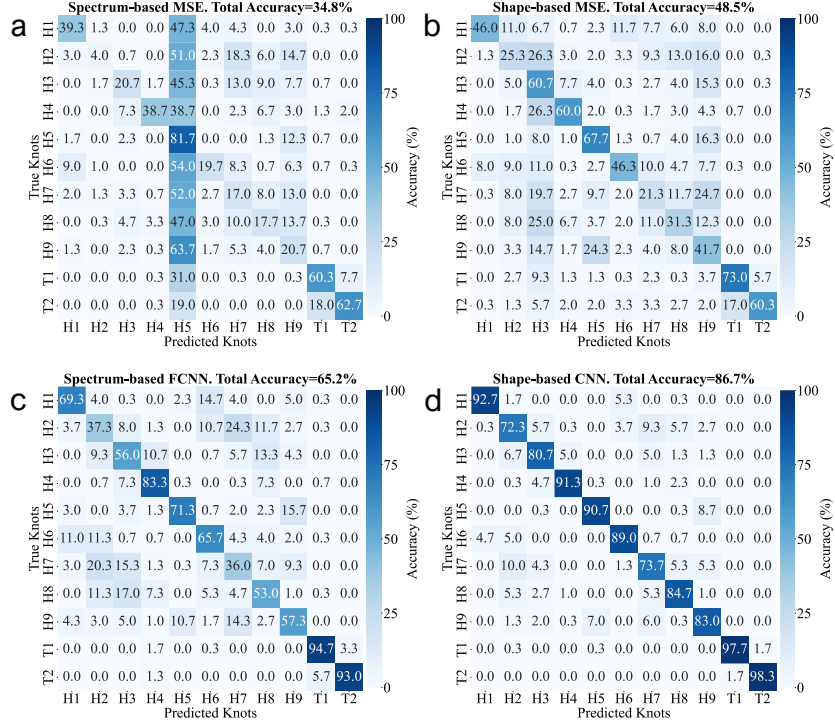

**Fig. 2. Confusion matrices for the classification of the Alphabet-11 knots after turbulence.** The confusion matrices display the percentage of each optical knot being classified into different knot types after experiencing turbulence. The "true knot" refers to the original knot type before turbulence (ground truth), while the "predicted knot" corresponds to the type assigned by the chosen classification method. The average classification accuracy is indicated at the top of each panel. Each confusion matrix incorporates data from all three turbulence strengths $\sigma_R^2 = 0.05, 0.15, 0.25$. (a) Spectrum-based MSE classification. (b) Shape-based MSE classification. (c) Spectrum-based FCNN machine learning classification. (d) Shape-based CNN machine learning classification.

The selected ranges of $l$ and $p$ values capture more than 99% of the total energy of alphabet-11 knots across $\sigma_R^2 = 0.05, 0.15$ turbulence cases. Under the strongest turbulence conditions, the total energy within the selected mode range may drop below 99% in some instances, but in most cases, it remains above 95%. This ensures that nearly all relevant spectral information is taken into account. Expanding the range further would offer minimal additional insight while significantly increasing experimental complexity, as measuring an extended set of LG modes with high precision is a significant challenge in laboratory experiments. Among the two MSE approaches, the filtered-spectrum MSE method yields higher classification accuracy, and therefore, in the main text, we refer to this approach when discussing MSE results. The results of the full-spectrum MSE approach are provided in Supplementary Materials S3.

The confusion matrix summarizing the performance of the MSE classification method, which combines all three turbulence strengths, is shown in Fig. 2a. Each turbulence strength and each knot type are represented by 100 samples, resulting in a total of 3,300 classified knots. The overall accuracy of the MSE-based spectral classification across all turbulence strengths is 34.8%, meaning that 34.8% of post-turbulence optical knots were correctly matched to their pre-turbulence LG spectra. The detailed performance breakdown for each turbulence strength is provided in Supplementary Materials S4. The spectral classification accuracy (34.8%) and the topological recovery rate (15.3%) are not directly comparable, since they measure fundamentally different aspects of turbulence-induced degradation. Spectrum-based classification evaluates how reliably the LG modal composition of a knot can be identified after turbulence, whereas topological recovery requires reconstructing the full 3D singularity line topology and assigning the correct topological class (e.g., Hopf link or Trefoil knot). Consequently, even when spectral information is partially preserved, the knotted topology may already be destroyed. Nevertheless, these results highlight the potential of mode multiplexing as an alternative transmission strategy for optical knots in turbulence.

**Mean Square Error Analysis of Singularity Line Shapes of Alphabet-11**

Here, we introduce a new approach that leverages the geometric stability of vortex line shapes. This method is motivated by the observation that, despite turbulence-induced distortions in LG spectra and topological features, the overall 3D structure of vortex singularity lines often remains visually similar before and after the turbulence. This effect is illustrated in Figs. 1a and 1 b, and additional examples are provided in Supplementary Materials S2. Similar observations have also been noted in our previous work [34], reinforcing the idea that vortex line geometry may provide a more resilient basis for classification compared to measures based on the number of crossings. This shape-based approach offers a fundamentally different perspective on structured light classification, providing a potentially more robust method for tracing knots in turbulent environments.

To test this approach, we implement an MSE-based shape classification method, where the post-turbulence 3D singularity shape of each knot is compared to its original shape. The knot volume is discretized into a 32×32×32 grid along the $x$, $y$, and $z$ axes, where each voxel is assigned a value of 1 if a vortex line crosses it and 0 otherwise. This resolution was chosen as it provides sufficient detail to capture the essential geometric features of vortex lines while remaining computationally efficient for MSE calculations, machine learning training, and visual analysis.. The MSE between the original shapes in Alphabet-11 and their post-turbulence counterparts is then computed, and the knot is classified based on the lowest MSE value. The classification results using such a shape-based MSE approach are presented in Fig. 2b. The overall classification accuracy for the shape-based MSE method is 48.5%, which is significantly higher than the 34.8% accuracy of the spectral MSE approach (for each turbulence strength, results see Supplementary Materials S4). This result demonstrates that vortex line shapes exhibit a higher degree of stability in turbulence compared to their LG spectra. Moreover, when compared to the 15.3% success rate of topological knot recovery, shape-based classification achieves an improvement of more than threefold, reinforcing the idea that optical knot shapes are a more robust measure of stability than their topological properties.

**Deep Neural Network Approach for Alphabet-11 Stability**

To further assess the effectiveness of shape-based classification, we extend our analysis beyond MSE-based methods and apply DNNs for optical knot identification. While the core concept, based on classifying post-turbulence optical knots according to either their LG spectra or 3D shapes, remains unchanged, we now employ machine learning to enhance classification accuracy and evaluate the performance of both approaches. First, we implement a fully connected neural network (FCNN) to classify Alphabet-11 knots based on their LG spectra, referring to this method as the spectrum-based FCNN approach. Details of the network architecture and training procedure are provided in the Supplementary Materials (S5). It is important to emphasize that our goal is not solely to maximize classification accuracy, but instead to (i) demonstrate the viability of LG mode-based classification as an alternative to topology-based approaches and (ii) systematically compare the two new approaches, being LG spectrum-based classification and shape-based classification, introduced in this study. The classification results for the spectrum-based FCNN are shown in Fig. 2c, demonstrating an overall accuracy of 65.2%. To compute this accuracy, we tested the network using 100 samples for each knot type ("class") in the alphabet at each turbulence strength, while 1000 additional samples were used for training. Accuracy values for individual turbulence strengths are provided in Supplementary Materials S6. This result demonstrates that even a simple neural network nearly doubles classification accuracy compared to the MSE-based spectral approach (34.8%), further reinforcing the potential of mode-multiplexed information in optical knots.

Next, we investigate whether shape-based stability remains superior to spectral classification when employing machine learning techniques. To test this, we implement a shape-based 3D convolutional neural network (shape-based CNN) trained on the same 32×32×32 voxel representation of singularity line structures used in the MSE-based classification. For a direct comparison, we again use 1000 samples for training and 100 samples for testing per element in Alphabet-11 for each turbulence strength. Details of the algorithm structure and training procedures are provided in the Supplementary Materials (S7). The classification results, averaged across all turbulence strengths, are shown in Fig. 2d, with individual confusion matrices for each turbulence level provided in the Supplementary Materials (S6). The overall accuracy for shape-based CNN classification reaches 87.6%, significantly surpassing the 65.2% accuracy of the spectral FCNN approach.

It may be argued that comparing spectral-based classification using a fully connected neural network (FCNN) with shape-based classification employing a convolutional neural network (CNN) is not entirely equitable, given the fundamental differences in network architecture. To address this concern, we also trained a spectrum-based 2D CNN model (see Supplementary Materials S7), achieving an accuracy of 62.6%, which closely matches the performance of the spectrum-based FCNN. Further details are provided in Supplementary Materials S8. The fact that both 2D CNN and FCNN models produce similar results reinforces the argument that these models are approaching their performance limits for mode spectrum-based classification. Meanwhile, the 3D CNN model trained on shape-based data outperforms both results by a substantial margin, highlighting the superior resilience of singularity lines' shapes to turbulence. Thus, our findings strongly suggest that the 3D shape of singularity lines of the optical knots is more robust to turbulence than their spectral mode content. This result indicates that a shape-based approach may offer a more stable and reliable framework for structured light applications, outperforming both topology-based methods and mode-multiplexed LG spectra-based approaches in turbulent environments.

**Flower Beams: A Shape-Based Transmission Approach**

Shape-based characterization of optical singularity lines presents a promising alternative to traditional topological and spectral approaches in structured light applications. To explore this concept further, we introduce Flower Beams, a novel class of structured optical fields with singularity lines shaped in a flower pattern, and demonstrate the resilience of their singularity shape in perturbative conditions, despite the absence of established topological stability properties. The concept of Flower Beams is illustrated in Figs. 3a and 3b. In what follows, we assign each petal in a Flower Beam a number corresponding to its relative size: full-sized petal → represented by 2; half-sized petal → represented by 1; missing petal → represented by 0. For instance, a Flower Beam with four fully sized petals is denoted as 2222 (shown in Fig. 3a), while a beam with one half-sized petal and one missing petal is denoted as 2120 (shown in Fig. 3c). For the detailed derivation of the mathematical expressions for the Flower Beams, see Supplementary Materials S9.

Using the above numbering convention scheme, a four-petal Flower Beam allows for $3^4 = 81$ distinct configurations, forming a new optical alphabet that we refer to as Alphabet-81. This framework is highly scalable – increasing the number of petals or introducing additional size petals (e.g., quarter-size) levels significantly expands the number of degrees of freedom. For example, using six petals with four size levels results in $4^6 = 4096$ unique configurations. Additionally, other degrees of freedom, such as petal rotation around their axis ($z$-axis), could further expand this concept [40]. Here, we consider beams with four petals, though examples with five and six petals are provided in Supplementary Materials S10.

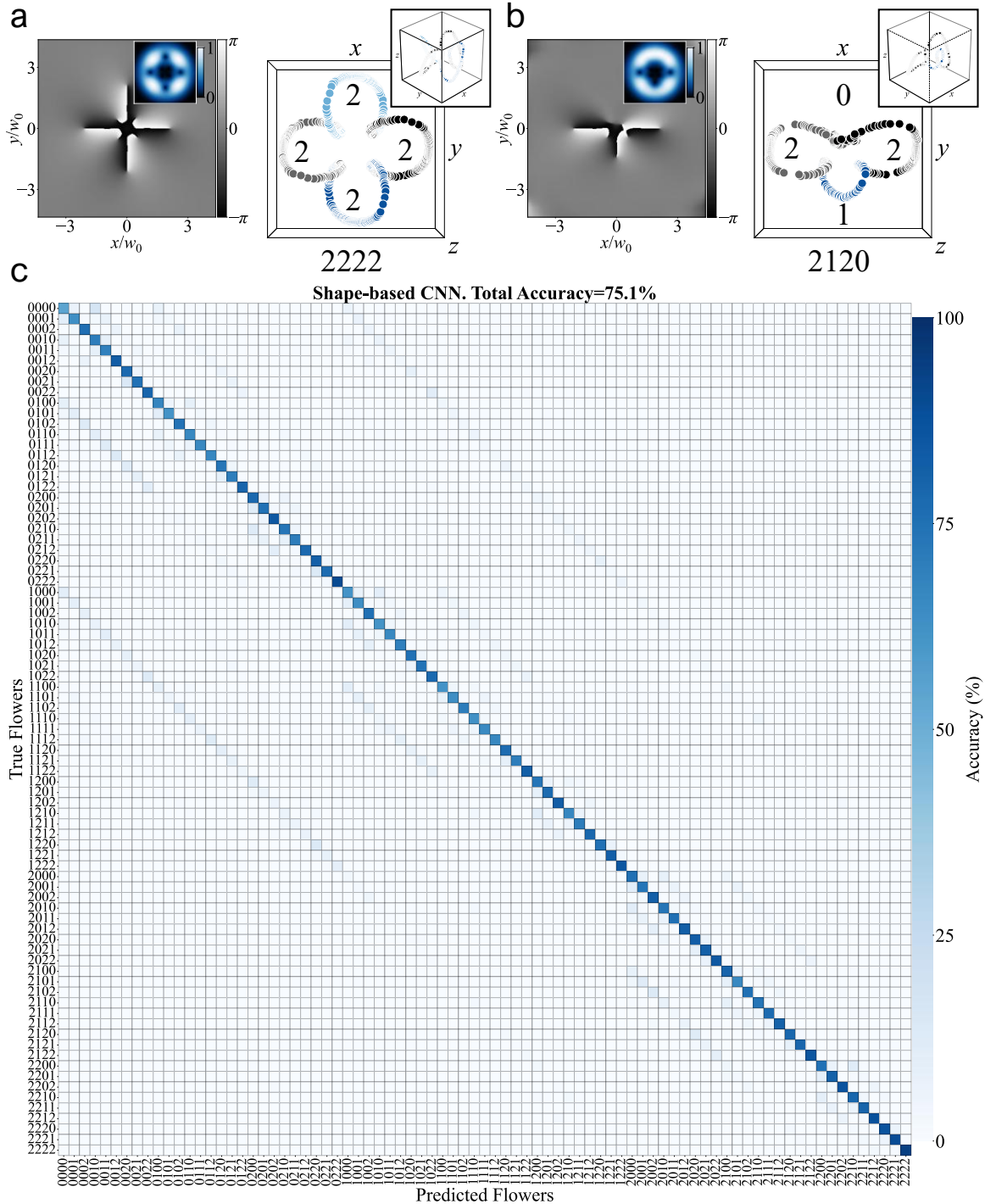

**Fig. 3. Petal size manipulation in Flower beams and their classification under turbulence.** Panel (a) presents a 2222 Flower Beam, where all four petals are full-sized. The left sub-panel displays the phase distribution of the field in the $z = 0$ plane, with an inset showing the corresponding amplitude distribution. The right sub-panels illustrate the 3D singularity structure from two perspectives: a general 3D view and a top-down view (inset). The petals are color-coded for better visual distinction. Panel (b) depicts a 2120 Flower Beam, where two petals are full-sized (2), one is half-sized (1), and one is removed (0). The layout follows the same structure as panel (a), facilitating a direct comparison of different petal configurations. Panel (c) shows the confusion matrix for the classification accuracy of various Flower Beam configurations after exposure to turbulence. In the confusion matrix, the true class corresponds to the original Flower Beam configuration before turbulence, while the predicted class denotes

the configuration assigned by the classification method. The results incorporate data from three turbulence strengths ($\sigma_R^2 = 0.05, 0.15, 0.25$), highlighting how well different petal structures are identified despite perturbations.

**Machine Learning Classification of Flower Beams in Turbulence**

To assess the stability of Flower Beams, we employ the same CNN-based classification approach used for Alphabet-11 optical knots. Each beam configuration is subjected to three different turbulence strengths $\sigma_R^2 = 0.05, 0.15$, and $0.25$ (see Supplementary Materials S11 for examples of Flower Beams after propagation in the media with varying strengths of turbulence), and we analyze how well their singularity line structures can be identified after propagation through turbulence. We use the same CNN architecture, modifying only the final layer to accommodate the larger number of classified classes (81 instead of 11). Further details about the network architecture and training procedures are provided in Supplementary Materials S7. For training, we generate 450 samples per Flower type per turbulence strength, with an additional 50 samples per class reserved for testing. The confusion matrix for classification across all three turbulence strengths is shown in Fig. 3c (see Supplementary Materials S12 for detailed confusion matrices for each turbulence level). Our results show that Flower Beams achieve high classification accuracy, with an overall success rate of 75.1% across all turbulence conditions. In weak turbulence, classification accuracy reaches 92.7%. A closer analysis of the confusion matrix reveals that misclassification primarily occurs between Flower Beams that differ by a single half-sized petal. Specifically, the most frequent errors arise when beams without half-sized petals are confused with those containing a single half-sized petal at the same position, as well as when half-sized petals are misclassified as full-sized petals. These patterns suggest that minor shape variations in petal size introduce subtle differences that challenge classification, particularly under stronger turbulence. This observation suggests that classification robustness could be further improved through targeted machine learning techniques. One potential approach is to implement a contrastive learning framework or modified loss function, which encourages the neural network to focus on distinguishing between visually similar classes by emphasizing the most relevant shape features. Additionally, optimizing the design of the Flower Beams themselves, such as refining petal size distributions or incorporating controlled petal rotations, could further enhance their stability and distinguishability in turbulent conditions. While this study focuses on fixed petal sizes as a proof of concept, future work could explore additional shape optimizations that maximize stability in turbulence, potentially leading to even higher classification accuracy and enhanced resilience for real-world optical applications.

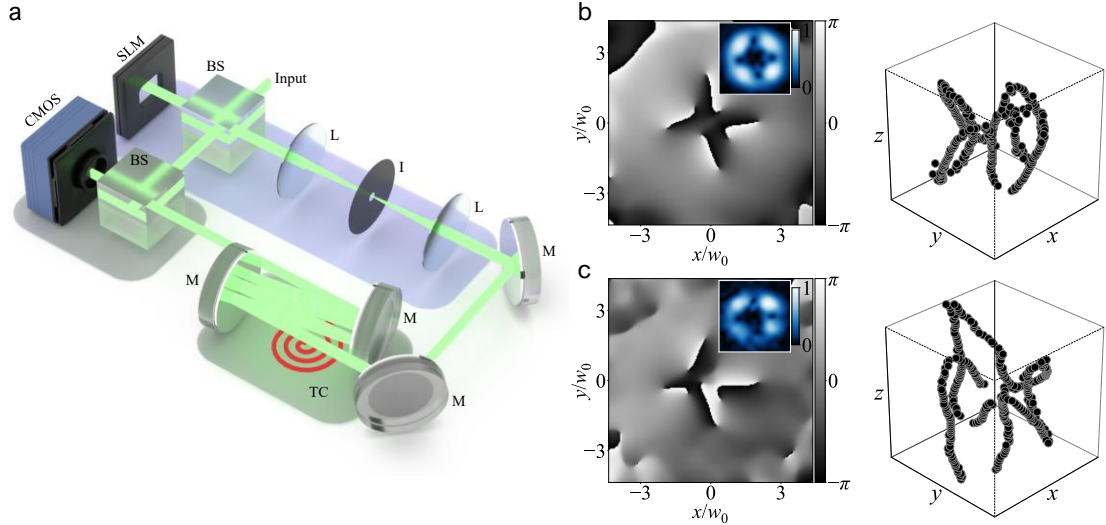

**Fig. 4. Experimental setup and measurement results.** (a) Schematic of the experimental setup used for generating and analyzing optical Flower beams. Here, BS = Beam Splitter, L = Lens, I = Iris, M = Mirror, TC = Turbulence Chamber, CMOS = Complementary Metal Oxide Semiconductor camera, SLM = Spatial Light Modulator. (b) Measured field phase and reconstructed 3D singularity structure of an unperturbed 2222 Flower beam at $z = 0$, with the inset showing the corresponding amplitude distribution. (c) The same measurements as in (b) but for the Flower beam after experiencing turbulence of strength $\sigma_R^2 = 0.05$, highlighting the perturbation effects on both phase and amplitude.

**Experimental Validation of Flower Beam Stability**

To validate the stability of Flower Beams under turbulent conditions, we conducted an experimental study using a controlled turbulence chamber. The experimental setup is based on a spatial light modulator (SLM) to encode the structured light field and a Mach-Zehnder interferometer [63], enabling a single-shot measurement of the complex electric field to recover singularity lines [37]. This method allows us to analyze how Flower Beams deform in realistic perturbative environments. The experimental setup is shown in Fig. 4a. The SLM is placed in the signal arm of the interferometer and is programmed to generate structured optical fields at a wavelength of 532 nm. The holograms displayed on the SLM correspond to the specific superposition of LG modes required to create each Flower Beam configuration. The structured beam propagates through a hot-air turbulence chamber, which simulates realistic atmospheric distortions by introducing controlled variations in refractive index [64]. The chamber is positioned at the image plane of the SLM, ensuring that the beam experiences turbulence as it propagates through the medium.

To optimize the experimental conditions, we scale the system parameters using the Fresnel number approximation. The Fresnel scaling parameter is given by $F = w_0^2/\lambda L$, where $w_0$ is the beam waist radius, $\lambda$ is the wavelength, and $L$ is the turbulence link length. This scaling enables us to simulate the propagation of a 6mm beam over a 270-meter turbulence channel in an experimental chamber, with a total propagation distance of 1.5m. The chamber utilizes heated air and convective airflow to generate controlled turbulence, enabling the study of its effects on structured beams. To ensure consistency between the turbulence levels in our numerical simulations and experimental conditions, we calibrated our laboratory turbulence chamber using heated airflow. This calibration involved measuring the variance of the beam wander displacement on the detector, which quantifies the intensity fluctuations caused by turbulence-induced variations in the refractive index. Comprehensive details of the calibration process, along with measurements of the refractive index structure parameter $C_n^2$ and the Fried parameter $r_0$, are provided in Supplementary Materials S13. In our experiments, a CMOS camera was used to capture the optical field at the waist plane, followed by the reconstruction of the singularity structures of the Flower Beams. An iris filters select the first diffraction order, ensuring that only the desired structured mode reaches the detector. The phase distribution $\varphi$ and amplitude $A$ are calculated numerically by inverting the expressions $\tan(\varphi) = (I_3 - I_2)/(I_1 - I_2)$ and $A = \sqrt{(I_3 - I_2)^2 + (I_1 - I_2)^2}$, where $I_i$ ($i = 1,2,3$) are phase-shifted interferograms with shifts of $(2i - 1)\pi/4$ [65].

We then applied the neural network model trained on simulation data to classify the experimentally generated Flower Beams after they were exposed to turbulence. However, due to experimental noise and real-world imperfections, we fine-tuned the model using a subset of experimental samples. The fine-tuning dataset consists of 10 samples per Flower Beam configuration at each turbulence strength. The classification accuracy as a function of the number of fine-tuning samples is presented in the Supplementary Materials (S14). The final classification accuracy for the experimental dataset is summarized in Table 1, alongside the results obtained from purely simulated data. In weak turbulence, the experimentally fine-tuned model achieved an accuracy of 91.2%, which is comparable to the 92.7% accuracy obtained in simulation, well within the standard deviation range. This confirms that the shape stability of Flower Beams observed in simulations is reproducible in real-world conditions. As the turbulence strength increases, classification accuracy degrades more rapidly in experiments than in simulations. Under the strongest turbulence $\sigma_R^2 = 0.25$, the experimental accuracy drops to 34.6%, whereas the model trained purely on simulated data achieves 59%. This discrepancy suggests that measurement imperfections – especially at higher turbulence levels – may contribute to additional distortions not accounted for in the simulations. Similar trends have been observed in our previous optical knot recovery experiments [34], where strong turbulence conditions led to significant measurement deviations due to experimental limitations. Several factors may contribute to the observed reduction in accuracy at higher turbulence strengths: (i) Temporal Variations in Turbulence: The experimental chamber introduces dynamic refractive index fluctuations, which lead to phase inconsistencies between different interference frames captured by the CMOS camera required for the phase recovery of the beam [34]. (ii) Limited Fine-Tuning Data: The model was fine-tuned using 10 samples per configuration, which may be insufficient for robust generalization in highly turbulent conditions. Increasing the fine-tuning dataset could potentially improve performance. (iii) Measurement Constraints on Higher Turbulence Levels: higher turbulence regimes lead to increased distortions, making experimental recovery more challenging, which likely impacts Flower Beam classification.

Our experimental results confirm the viability of Flower Beams as a robust method for structured light transmission. The high classification accuracy in weaker turbulence regimes supports the conclusion that shape-based transmission outperforms traditional topological and spectral methods. The experimental validation further reinforces the idea that geometric stability of singularity structures is a promising alternative approach for structured light applications.

**Discussion**

In this study, we addressed a critical challenge in structured optical transmission, the degradation of singular optical beams due to turbulence-induced perturbations. Although traditional approaches rely on topological invariants for robust transmission, our analysis revealed fundamental limitations of this approach. Despite their theoretical robustness, optical knots undergo reconnection events in turbulence, which significantly diminishes their practical reliability in applications such as communication and sensing. To overcome these constraints, we introduced and validated an innovative approach emphasizing the unprecedented stability of optical singularity line shapes. Initially, we demonstrated through extensive numerical simulations and analysis of optical knots (such as the trefoil knot and Hopf link structures) that shape stability consistently outperformed traditional spectral and topology-based methods under various turbulent conditions. Building upon these insights, we introduced Flower Beams that are structured optical fields designed explicitly for robust transmission based on geometric singularity shapes. Our deep learning-driven analysis of Flower Beam trajectories showed a substantial improvement in classification accuracy compared to conventional methods. Specifically, shape-based classification using convolutional neural networks achieved accuracies exceeding 90% in weak turbulence and remained robust in stronger turbulence cases, highlighting the potential of shape-based transmission for practical turbulent propagation scenarios. Experimental validations confirmed the applicability and robustness of Flower Beams. Despite real-world limitations, including measurement inaccuracies and dynamic temporal fluctuations inherent in experimental setups, our findings consistently agreed with numerical predictions. This consistency highlights the fundamental stability advantage of shape-based methods over traditional topological and spectral approaches.

Looking ahead, several strategies could further enhance the resilience of shape-based transmission. Introducing additional degrees of freedom, such as varying petal rotations, incorporating asymmetrical configurations, or adjusting petal sizes, could significantly improve discriminability and resistance to severe turbulence. Furthermore, applying advanced machine learning techniques, such as contrastive learning or optimized neural network architectures, may further boost classification performance, especially under challenging conditions. Overall, by establishing singularity line shape stability as a robust and scalable method, this work opens new opportunities for enhanced free-space optical communication, quantum information processing, and advanced imaging technologies. Future research can build upon these findings to develop novel, turbulence-resilient optical systems, thereby expanding their applicability in both classical and quantum optical technologies.

| Table 1. Classification Accuracy Under Varying Turbulence Levels. | | | | |
|---|---|---|---|---|
| **Rytov Variance $\sigma_R^2$** | **0.05** | **0.15** | **0.25** | **0.05 + 0.15 + 0.25** |
| **MSE-Based LG Spectrum Classification (Alphabet-11)** | 50.3±3.0 | 32.4±2.8 | 21.6±2.4 | 34.8±1.6 |
| **MSE-Based Singularity Shape Classification (Alphabet-11)** | 72.1±2.7 | 42.8±2.9 | 30.6±2.7 | 48.5±1.7 |
| **CNN-Based LG Spectrum Classification (Alphabet-11)** | 79.6±2.4 | 59.1±2.9 | 49.1±3.0 | 62.6±1.7 |
| **FCNN-Based LG Spectrum Classification (Alphabet-11)** | 79.7±2.5 | 62.0±2.9 | 53.8±2.9 | 65.2±1.6 |
| **3D CNN-Based Singularity Shape Classification (Alphabet-11)** | 97.7±0.9 | 87.1±2.0 | 75.4±2.5 | 86.7±1.2 |
| **CNN-Based Flower Beam Classification – Simulations (Alphabet-81)** | 92.7±0.8 | 73.6±1.4 | 59.0±1.5 | 75.1±0.8 |
| **CNN-Based Flower Beam Classification – Experimental Data** | 91.2±2.0 | 60.7±3.4 | 34.6±3.3 | 62.2±1.9 |

## Materials and methods

### Turbulence Simulations

In this study, we simulated atmospheric turbulence by generating phase distortions using multiple phase screens and using the angular spectrum method for numerical beam propagation [37]. For the generation of phase screens, we employed the von Kármán turbulence model with subharmonic augmentations [66], implemented using the Python package for AO tools [67]. The turbulence model was characterized by an outer scale, $L0$, of 10 meters and an inner scale, $l0$, of 3 mm [66, 68, 69]. To accurately represent turbulence in simulations, we segmented the total propagation distance into multiple sections, each approximating weak turbulence conditions. The number of phase screens required, $N_{sc}$, was determined using the relationship $N_{sc} = (10\sigma_R^2)^{-6/11}$, where the Rytov variance $\sigma_R^2$ is defined as $\sigma_R^2 = 1.23 C_n^2 k^{7/6} L^{11/6}$, with $C_n^2$ being the refractive index structure parameter, $k = 2\pi/\lambda$ the optical wavenumber, and $L$ the total propagation distance [70]. For our setup with $L = 270$m, the beam size of 6mm ($1/e^2$), and wavelength $\lambda = 532$nm [34,71], this resulted in $N_{sc} = 3$ being sufficient for each of the turbulence conditions tested ($\sigma_R^2 = 0.05, 0.15, 0.25$). The corresponding refractive index structure parameters $C_n^2$ for these turbulence strengths were set to $8.0 \times 10^{-15}$, $2.4 \times 10^{-14}$, and $4.0 \times 10^{-14}$ m$^{-2/3}$, respectively. The Fried parameter, given by $r_0 = 1.68(C_n^2 L k^2)^{-3/5}$, yielded values of $5.47, 2.83$, and $2.08$cm for the respective turbulence strengths.

### Acknowledgements

This work was supported by the Multidisciplinary University Research Initiative (N00014-20-2558), the Office of Naval Research (N00014-25-1-2402), and the Army Research Office (W911NF2310057).

### Contributions

D.T., D.G.P., and N.M.L. conceived the idea for this study and performed theoretical and numerical analyses. D.T. developed the concept of singularity-shape-based stability, performed numerical studies, and deep-learning-based analysis of knot stability. D.G.P. built the experimental setup and performed the experiments. All authors contributed to the discussion of the results and writing of the manuscript.

### Data availability statement

The data is available from the authors on reasonable request.

### Conflict of interest

The authors declare no competing interests.